# Using Integer Constraint Solving in Reuse Based Requirements Engineering

Camille Salinesi, Raul Mazo, Daniel Diaz, Olfa Djebbi
Centre de Recherche en Informatique
Université Paris 1 Pantéhon – Sorbonne
Paris, France
Camille.Salinesi@univ-paris1.fr, Raul.Mazo@malix.univ-paris1.fr,
Daniel.Diaz@univ-paris1.fr, Olfa.Djebbi@malix.univ-paris1.fr

*Abstract*—Product Lines (PL) have proved an effective approach to reuse-based systems development. Several modeling languages were proposed so far to specify PL. Although they can be very different, these languages show two common features: they emphasize (a) variability, and (b) the specification of constraints to define acceptable configurations. It is now widely acknowledged that configuring a product can be considered as a constraint satisfaction problem. It is thus natural to consider constraint programming as a first choice candidate to specify constraints on PL. For instance, the different constraints that can be specified using the FODA language can easily be expressed using boolean constraints, which enables automated calculation and configuration using a SAT solver. But constraint programming proposes other domains than the boolean domain: for instance integers, real, or sets. The integer domain was, for instance, proposed by Benavides to specify constraints on feature attributes. This paper proposes to further explore the use of integer constraint programming to specify PL constraints. The approach was implemented in a prototype tool. Its use in a real case showed that constraint programming encompasses different PL modeling languages (such as FORE, OVM, or else), and allows specifying complex constraints that are difficult to specify with these languages.

## I. Introduction

Product Line (PL) engineering has become an unavoidable approach to support reuse in systems development. The PL approach helps realize order-of-magnitude improvements in time to market, cost, productivity, quality and flexibility. In research, many works were devoted to defining PL modeling languages, and others to PL configuration (ie defining product that reuse assets as defined in the PL models). Several works have shown that constraints play a central role in the determination of which products are permitted and which are not.

This starts with the FODA notation, which offers ways to -starting from a bundle of features- constraints the number of those that can be included in a configuration. The "excludes" and "requires" constraints also allow us to specify that when a product includes a feature then, another one should be excluded or included too.

Van Deursen [1] proposed to reason on feature models by translating them into a logic program using predicates such as all( ), one-of( ), or more-of( ), that respectively specify mandatory, mutually exclusive, and alternative features. For instance constraints:
  F1 = all (F2, F3, F4)
  F4 = one-of (F5, F6)
specify that if F1 is included in a configuration, then F2, F3, and F4, and therefore either F5 or F6 should be included too.

The use of constraint programming to reason about feature model was extended by Batory [2], who proposed an approach to transform a feature model into propositional formulae using the ∧, ∨, ¬, ⇒ and ⇔ operations of propositional logic. This enables for example constraints of the form
  F => A ∨ B ∨ C
  meaning that feature F needs features A or B or C, or any combination thereof. As Van Deursen's [1] and Mannion's approaches [3], in these constraints, features are boolean variables (either they are included or not in a configuration). It is then possible to use a SAT solver to ensure the satisfiability of the set of boolean formulae. Another approach consists in using Constraint Programming (CP). Indeed, CP is a powerful paradigm for solving combinatorial problems arising in many domains, such as scheduling, planning, vehicle routing, configuration, networks or bio-informatics. The idea of CP is to solve problems by stating constraints and finding a solution satisfying all the constraints. A constraint is simply a logical relation between several unknowns, these unknowns being variables that should take values in some specific domain of interest. A constraint thus restricts the degrees of freedom (possible values) the unknowns can take; it represents some partial information relating the objects of interest. The execution of a program mainly adds the constraints (incrementally) and asks the built-in solver to find a solution (an assignment of variables that satisfies the constraints). There are solvers for various domains: Finite Domains, Reals, Rationals, Booleans, Trees, Lists, Sets, Strings, etc. Among these domains, Finite Domain (FD) is the most useful in practice. An FD variable can take values inside an initial domain composed of a finite set of integers. An FD solver uses consistency techniques borrowed from CSP to maintain the consistency of the constraints. Obviously, it is possible to use an FD solver to solve boolean constraints. In [4] we have shown how to use FD constraints on [0..1] variables to efficiently encode boolean constraints such as ∧, ∨, ¬. We have also shown that for many problems, an FD solver can outperforms specific boolean solvers (SAT, BDD-based, 0-1 programming). The initial work of Benavides used this approach to ensure the satisfiability of the boolean formulation associated to a PL model.

Benavides et al [5] extended their previous work to reason about constraints specified on feature attributes also modeled as FD variables. Constraints such as

F1.A = F2.B + F3.C

can be specified to express that in any configuration, the value of attribute A associated with feature F1 should be equal to B+C where B and C are attributes respectively associated to F2 and F3. This allows to reason on extra functional features as defined by Czarnecki [6], ie relations between one or more attributes of one or different features.

In this paper we propose to go further and to exploit more deeply the richness of CP over FD. We implemented these ideas in an interactive tool that allows the user to define a model (using various meta-models: FODA, FORE, OVM, and MAP), to configure it (possibly adding extra-constraints), to explore various solutions, to backtrack and change some settings before a new derivation. This tool is based on our GNU Prolog system [7], which contains an efficient constraint solver over FD. Such a solver offers a wide variety of constraints, which we think have not yet been exploited to their full potential. For instance GNU Prolog offers:

- arithmetic constraints (both linear and non-linear), e.g. X+Y < Z or X*Y<>Z. The use of min and max is also allowed inside those constraints.
- symbolic constraints, e.g. atmost (2,[X,Y,Z,T],10) states that at most 2 variables among X,Y,Z,T can take the value 10. As another example the symbolic constraint element(I, [v1,v2,…,vN],X) enforces the variable X to be equal to the Ith element of the vector of N values [v1,…,vN].
- boolean constraints: GNU Prolog offers all boolean constraints such as ∧, ∨, ¬, ⇒, ⇔,… Variables appearing in such constraints are implicitly constrained to the domain [0..1].
- reified constraints: making it possible to reason on the issue (unsatisfied /satisfied) of a constraint. Namely, a constraint *C* can appear inside any (above) boolean constraint (constraints are first-class objects). As an example consider the boolean constraint X<Y => K=8. Its operational behavior is : as soon as the solver detects that X<Y it enforces K=8, conversely if it discovers K<> 8 it enforces X>=Y.

Constraint programming has already been explored before to support the specification and analysis of PL. We believe that our approach is original because (a) it supports the specification of constraints that today can only be specified with various languages, (b) it supports in an integrated way the analysis of PL model constraints that so far can only be analyzed with separate approaches, and (c) it supports the specification of new kinds of constraints both on PL models and product requirements.

For example, our approach allows us to implement reified constraints in a FODA PL model, such as: *"whenever a feature F1 is included in a product, then constraint C (e.g. F2 excludes F3) shall be enabled"*.

This enables dynamic configuration by expressing extra constraints at configuration time. This is particularly important in practice since it is necessary to add/relax some constraints for given derivations [8]. Reified constraints can also be used to implement OVM's variation point dependencies as proposed by Pohl et al [9].

Another interesting aspect of our approach is that the GNU-Prolog solver that is used supports the analysis of any kind of finite domain constraints, such as: *"the value of attribute F1.A should always be equal to F2.B + F3.C"* to control the value of integer feature attributes, as proposed by [5], but at the same time it permits to control the number of occurrences of a feature, as for instance in the constraint *"a product should include at least 2 and at most 4 occurrences of feature F"*.Feature cardinalities were proposed by Czarnecki [6], but constraint analysis on feature cardinalities has not yet been tackled to our knowledge [10], and there is no tool available so far to support the analysis of constraints on feature cardinalities and on feature attributes in an integrated way. Finite domain constraints can also apply on any ENUM PL properties, like in the Decision King tool which uses them to control decision consistency [11].

Our approach also enables the specification of "complex" product requirements (complex compared to select or not a feature) under the form of additional constraints specified at the moment of configuration. For examples, our approach supports the specification of constraints such as *"provide me with all possible configurations in which the value of feature attributes A1..Ai is in [a..b]"*. This is useful in staged configuration [12]. Other new kinds of product-specific constraints such as: *" provide me with a configuration in which the values of all the attributes associated with features F1..Fn are different from each other"*, and *"provide me with all product configurations in which features F1...Fn are either all included or all excluded"*. Such constraint can be used to query the PL model, that is useful for instance to explore configuration scenarios, or in a verification activity.

Last, constraint programming is efficient in solving optimization problems. Our approach supports the specification and analysis of goals such as *"identify the optimal configuration with respect to cost (min goal) and benefit (max goal) feature attributes"* to detect "optimal" products and support decision making during the configuration activity [13].

The rest of the paper is structured as follows: section 2 presents a working example, which is used in section 3 to illustrate a series of types of constraints over FD that we propose to support PL specification and analysis. Section 4 shows how these constraints are implemented in the GNU Prolog constraint reasoning platform. Section 5 discusses the effectiveness of our approach in the light of its application to a real example. Section 6 discusses related works. The concluding section presents our perspective on future works in the domain of constraint-programming based Product Line engineering.

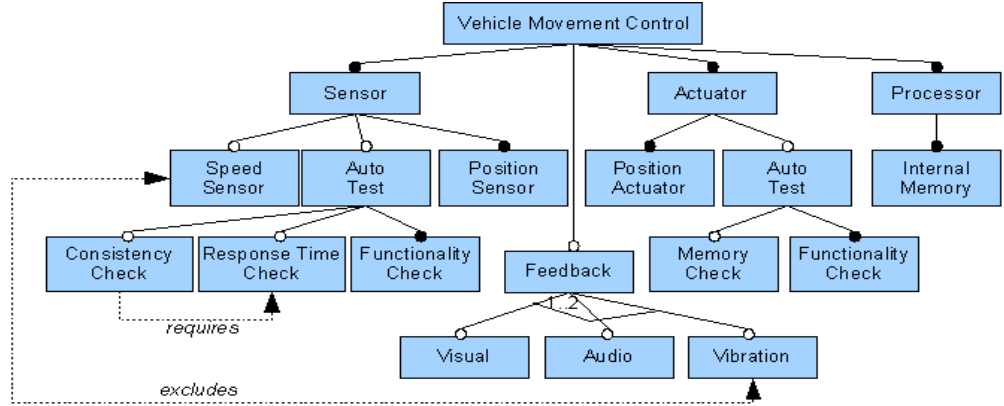

Figure 1. example of a simplistic vehicle movement control systems product line.

## II. WORKING EXAMPLE

Figure 1 shows the PL model of a family of simplified Vehicle Movement Control (VMC) systems using the FODA notation.

The figure shows that:
- control systems have four mandatory components, namely sensors, actuators and processors, and a fifth optional one for feedback;
- feedback can be visual, audio or by vibration, several kinds of feedback can be chosen, two at most in a single configuration;
- sensors can either detect position or speed, and they can have an auto-test;
- position sensors are mandatory;
- speed sensors are optional;
- when a speed sensor is included in a configuration, then vibration feedback must be excluded and conversely;
- actuators can have auto tests to check functionality (mandatory), and memory (optional);
- in addition to functionality, sensor auto-tests can check response time, and consistency check;
- when a consistency check is included in a configuration, then response time check must be included too;
- processors have internal memory.

## III. SPECIFYING PRODUCT LINE CONSTRAINTS USING FINITE DOMAIN

The following subsections illustrate various kinds of constraints in the FD from the most simple (and common with respect to state of the art), to the most complex (and original).

### A. Modeling FODA-like constraints

Specifying that a feature identified in a FODA model can be either included or excluded can very simply be done by defining a [0..1] domain to the feature where the 1 value would mean that the feature is included in the configuration, and the 0 value that it is not.

Traditional FODA constraints can be specified on the [0..1] domain as follows:
- F2 is a mandatory subfeature of F1 : *F2 = F1*;
- F2 is an optional subfeature of F1 : *F2 <= F1*;
- F1 requires F2 : *F1 <= F2*;

F1 excludes F2 can be specified as : *F1 + F2 <= 1*. Another specification could be *F1 * F2 = 0* (GNU Prolog accepts non-linear constraints, however in some cases non-linearity can penalize efficiency);
- Min-Max cardinality of a bundle of subfeatures F1..Fk of a feature F: *Min <= $\Sigma_{1..k} F_i$, and $\Sigma_{1..k} F_i$ <= Max*.

In the VMC example, the Feedback, Visual, Audio, and Vibration features are features that can only be either included or not in a product configuration. Besides, at least one, and at most 2 of these features can appear at the same time in the same configuration. The constraints are thus the following ones :

*Visual <= Feedback,*
*Audio <= Feedback,*
*Vibration <= Feedback,*
*Visual + Audio + Vibration >= 1,*
*Visual + Audio + Vibration <= 2.*

Constraint programming on the [0..1] domain permits to specify other kinds of less usual kinds of constraints, such as:
- *F1 + F2 > 0* : this is an equivalent to the logical (inclusive) OR (as proposed by Mannion [3]): F1 or F2 or both features can be included in a configuration. In GNU Prolog we can simply write F1 \/ F2.
- *F1 + F2 = 1* : this constraint specifies a logical (exclusive) XOR. It indicates that either F1 or F2, but at least one of them, should be integrated in the configuration. The XOR shall be distinguished from FODA's 'excludes' according to which a configuration can include none of the features.
- $\Sigma_{1..k} F_i > 0$ : this allows to specify that at least one of the Fi should be selected, ie equivalent to a min cardinality constraint applied to a collection of independent features.
- $\Sigma_{1..k} F_i = 1$ : this constraint specifies that only one of a collection of features can be selected.
- $\Sigma_{1..k} F_i = 0$ (alt < 1) : this constraint specifies that all the features of a collection are excluded. This constraint can be very useful in practice to impose some extra-restriction at configure-time.

Back to the VMC example, let us assume that in addition to the cardinality constraints, the visual and audio feedback features are mutually exclusive: any configuration shall include either the one or the other but at least one of them. This XOR dependency can be specified with the constraint:

$$Visual + Audio = 1$$

It is interesting to notice that whereas constraints on pairs of features could easily be represented graphically in a FODA model, constraints programming allows to specify constraints on collection of features that can be quite distant from each other. Representing graphically such kind of feature in a model would be difficult and rapidly entail readability.

Another interesting aspect of these constraints is that they could be rather straightforwardly be specified using boolean constraint programming. Specifying constraints in FD permits to specify other kinds of constraints that could more difficultly be specified with boolean constraints. For example:
- $F1 > F2$ : F2 is required and F1 excluded.
- $F1 + F2 <= F3$ : if F3 is included then either F1 or F2 is included; otherwise all are excluded.
- $F1 + F2 < F3 + F4$ : more features from the {F3, F4} set shall be included than from the {F1, F2} set. This feature can, of course, be extended to larger sets.
- $2 * F1 + F2 + F3 = 2$ : either F1 is included in the configuration, or both F2 and F3. Interestingly, one solution to specify this kind of constraint in FODA would be to artificially creating a fourth feature that represents F2 + F3. Of course the issue would then be of the actual meaning of this feature, of the link between F2 and F3 and their parent features, and of the place of the F4 feature in the model, which makes such a solution very unlikely.

B. *Reasoning about the number of occurrences of features*

In [Czanercki05], Czarnecki et al. introduce feature cardinality as the number of times a feature can be repeated in a product. This is useful to specify the number of times a feature is included in a product, as in bill of materials used in production, or to define a proportion, as defined in recipes in different sectors of the process industry, such as pharmaceutics, petro-chemistry, etc.

This can be modeled associating an FD variable F to a feature whose initial domain is 0..N (N being the maximum number of occurrences of F). Two constraints are obviously needed :
- $F1 > a$ : to indicate that feature F1 shall be included at least one time in a configuration, and
- $F1 = a$ : to specify the exact number of times a feature can be included in a configuration.
- *But more complex constraint can arise, for instance:*
- at least one of two features should be included in a product: $F1 + F2 > 0$
- at least one of a series of features should be included in a product: $\Sigma_{1..k} F_i > 0$
- there should be only one occurrence of two multivalued features: $F1 + F2 = 1$
- the product should include no occurrence of a series of features: $\Sigma_{1..k} F_i = 0$ (alt < 1)
- the product should include more occurrences of a feature than of another: $F1 > F2$
- the product should include more occurrences of a feature F1 than of two other features (F2 and F3) together: $F1 + F2 <= F3$; this is for example useful to specify that the VMC products should include more processors than sensors and actuators.
- the product should include more occurrences of a pair of features (F3, F4) than of another pair of features (F1, F2) together: $F1 + F2 < F3 + F4$; this is, for instance, useful to specify that the number of consistency check plus the number of response time auto test sensors should be superior to the number of memory check + the number of functionality checks in actuators
- the number of occurrences of F1 should be the half of the number of occurrences of F2: $2 * F1 = F2$; this can be used to specified that there should be two functionality checks auto tests per speed sensor.

The specification of requires and excludes relationships is a little different when applied to multi-occurrence features than with [0..1] features. For instance, in the VMC example, Vibration and Speed Sensor are mutually exclusive. This exclusion can be specified by the constraint:

*(SpeedSensor <> 0) <=> (Vibration = 0)*

or more simply

*(SpeedSensor > 0) XOR (Vibration > 0)*

The VMC example also contains a requires dependency from Consistency Check to Response Time Check. This can be specified by the constraint:

*ResponseTimeCheck <= ConsistencyCheck*

However, another kind of "requires" dependency could be specified to indicate that each instance of a feature requires an instance of another feature. In the VMC example, this would be the case if one actuator is required for each sensor. The constrain is specified as

*Actuator >= Sensor*

If in addition, actuators are only needed in product to control sensors, then the requires dependency is specified with an even stricter constraint:

*Actuator = Sensor*

or if n additional sensors are needed for other purposes:

*Actuator + n = Sensor*

Another use of constraints on the [0..n] domain in PL is to apply them to feature attributes, as proposed by Bennavides [5] with attributes associated to {true, false} features. In [13], we have demonstrated how to specify constraints on attributes to reason on goal based product configuration, to guide for example a cost/benefit analysis of products during their configuration.

### C. Reified constraints

In [12], Czarnecki et al. define a new feature modeling language to account for staged configuration. The fundamental of stage configuration is to enable constraints that shall be associated to a configuration model, which shall itself be considered as a PL model.

Staged configuration can be found useful when not all constraints shall be verified at once, but enabled in a ordered fashion. In general the reification of the constraint C into a variable B of the [0..1] domain is achieved by a constraint:

$$C <=> B = 1$$

that establishes a correspondence between B and *C* as follows: B = 1 iff *C* is true (thus B <> 1 (i.e. equal to 0) iff *C* is false).

The constraints of a PL model that shall only be verified at a stage of configuration identified must be reified. Identifying stages of configuration can be done either using an FD variable that represents time (the version number of the configuration), or it can be conditioned by the inclusion of a feature in the configuration. The following constraints shall be specified in the latter case:

- *F1 = 1 => B* : whenever F1 is included, the constraint *C* reified with the B variable should be satisfied.
- *F1 = 0 => B* : whenever F1 is excluded, the constraint *C* reified with B should be satisfied.

Of course, these constraints reification could also be directly specified as:

$$F1 = 1 => C, \text{ and}$$
$$F1 = 0 => C.$$

In the VMC example, it would for instance be possible to generate PLM models from the PL model to specify sub families of VMC. One interesting such kind of sub family is this in which a position actuator is associated with each position sensor. Another aspect of this sub-family is that it can be managed only as soon as there is a central processor with a 1024 Ko internal memory.

The constraint can be reified as follows :
$$(Processor = 1) \land (InternalMemory = 1024) => B$$
with
$$B <=> (Position\_Sensor = Position\_Actuator)$$

Reified constraint can also be used to specify constraint over decision points, as in [9] as follows. Assuming that a decision point D, specified using constraints C1..Cn, a constraint C on D shall simply be expressed as:

$$C => D, \text{ where}$$
$$D <=> C1 \land .. \land Cn$$

to indicate that whenever condition C is met (e.g. a feature is included in a configuration: F>0), the constraints associated with decision point D shall be satisfied.

### D. Symbolic constraints

CP over FD supports the specification and analysis of symbolic constraints, ie constraints that are checked on collections of variables. Here are some symbolic constraints:

- *Alldifferent(F1, .., Fk)* : specifies that in any configuration the value of each of the F1 to Fk features should be different pairwise. This could be specified by k (k-1) / 2 inequality constraints between each par of feature.
- *Atmost(n, F1..Fk, a)* : specifies that at most n of the F1 to Fk features are equal to a.
- *Atleast(n, F1..Fk, a)* : specifies that at least n of the F1 to Fk features are equal to a.
- *Exactly(n, F1..Fk, a)* : specifies that exacltyt n of the F1 to Fk features are equal to a.
- *Relation(F1..Fk, {a1..ak})* – constraints the tuple of Feature F1..Fk to be equal to at least one tuple in the collection of tuples {a1..ak}. This allows to specify extensively a predetermined collection of compatible values for [0..n] features.

In the VMC example, symbolic constraints can be used for instance to specify predefined combinations of the number of Sensors, actuators and internal memory in configurations:

*Relation ([Sensor, Actuator, InternalMemory], [*
*[1, 1, 32],*
*[1, 2, 64],*
*[2, 1, 64],*
*[2, 2, 128],*
*[3, 3, 512]*
*[4, 4, 1024]])*

Reified constraints could also be used to specify the $choose_{n,m}$ (F1, … ,Fk) predicate proposed by [2] to indicate that at most m and at least n of the F1...Fk {true, false} constraints shall be included (true) in the configuration. The specification can be done with two symbolic constraints :

*Atleast (n, F1..Fk, 1),*
*Atmost(m, F1..Fk, 1)*

The versatility of CP can also be used to specify these constraints on {true, false} feature using two generalized versions of the $\Sigma_{1..k} Fi > 0$ constraint:

$$min <= \Sigma_{1..k} Fi$$
$$\Sigma_{1..k} Fi <= max$$

which indicates that at least min and at most max features of the collection of Fi features shall be included in the same configuration.

One can easily see that these constraints offer a power of expression that go beyond the choice predicate. Indeed, they can not only be used to specify the minimal and maximal number of features to exclude, but also with [0...n] features to specify the minimal and maximal number of features that are instantiated a given number of times in the product.

## IV. TOOL IMPLEMENTATION

Developing a constraint program that specifies a product line model is quite straightforward. For example, the feature model presented in Figure 1 augmented with some of the

constraints presented in section 3 can be specified with the following program:

```
pl(L):-
L = [VMC, Sensor, Actuator, Processor,
     Feedback, SpeedSensor, SensorAutoTest,
     PositionSensor, PositionActuator,
     ActuatorAutoTest, InternalMemory,
     ConsistencyCheck, ResponseTimeCheck,
     SensorFunctionalityCheck, Feedback,
     MemoryCheck,
     ActuatorFunctionalityCheck, Visual,
     Audio, Vibration],
     fd_domain([VMC], 0, 1),
     fd_domain([Sensor, SpeedSensor,
     PositionSensor, SensorAutoTest,
     ConsistencyCheck, ResponseTimeCheck,
     SensorFunctionalityCheck], 0, 4),
     fd_domain([Actuator, PositionActuator,
     ActuatorAutoTest, MemoryCheck,
     ActuatorFunctionalityCheck], 0, 100),
     fd_domain([Feedback, Visual, Audio,
     Vibration], 0, 1),
     fd_domain([Processor], 0, 1),
     fd_domain([InternalMemory], [32, 64,
     256, 512, 1024]),
     Sensor #> VMC,
     Actuator #> VMC,
     Processor #> VMC,
     PositionSensor #= Sensor,
     SpeedSensor #=< Sensor,
     SensorAutoTest #=< Sensor,
     SensorFunctionalityCheck #=
     SensorAutoTest,
     ConsistencyCheck #=< SensorAutoTest,
     ResponseTime #=< SensorAutoTest,
     ResponseTime #=< ConsistencyCheck,
     Visual #=< Feedback,
     Audio #=< Feedback,
     Vibration #=< Feedback,
     Visual + Audio + Vibration #>= 1,
     Visual + Audio + Vibration #=< 2,
     PositionActuator #= Actuator,
     ActuatorAutoTest #=<Actuator,
     MemoryCheck #=< ActuatorAutoTest,
     ActuatorFunctionalityCheck #=
     ActuatorAutoTest,
     SpeedSensor #\= 0 #<=> Vibration #= 0,
     fd_labeling(L).     % To find one
     solution
```

We developed an interactive tool that :
- offers a graphical editor [14] that supports the drawing, loading and saving of PL models specified with different languages (FORE, XMI, and textual constraints); A screen dump of our tool interface is shown in Figure 2;
- compiles the model and generates the associated constraint program (as in the above example). This program can be consulted and/or modified by the user;
- allows the user to start a configuration process. This includes the addition of extra-constraints (whose lifetime is the configuration phase only), the addition of some optimization criterion, the computation of a solution, the interactive exploration of alternative solutions,…
- supports multi-language PL modeling and analysis tool [15].

A model validity checker is being implemented to support the analysis of a series of structural and semantic verification criteria [14]. So far, the verification function is implemented using procedural programs, whereas in fact our verification criteria were defined using first order logic., here again the underlying Prolog system will help us.

## V. FEASIBILITY STUDY WITH A REAL CASE STUDY

One particular question that can be raised about the new kinds of constraints that have been identified in this paper is "are they useful?". Although only long term experience shall provide a definitive answer to this question, one might be interested in looking for special constraints that could be specified in a real case.

To do so, we have used our CP over FD approach to specify constraints on a family of blood analysis automatons. We had the opportunity to model this PL using the FODA in the context of a cooperation with the STAGO company [13].

Applying the derivation rules proposed at the beginning of section 2.1 on the feature model shown in figure 2 generated about around 50 constraints on [0..1] features.

Of course, constraints that were associated to {true, false} features were, as we propose it, declared with a [0..1] domain to enable the addition of supplementary constraints.

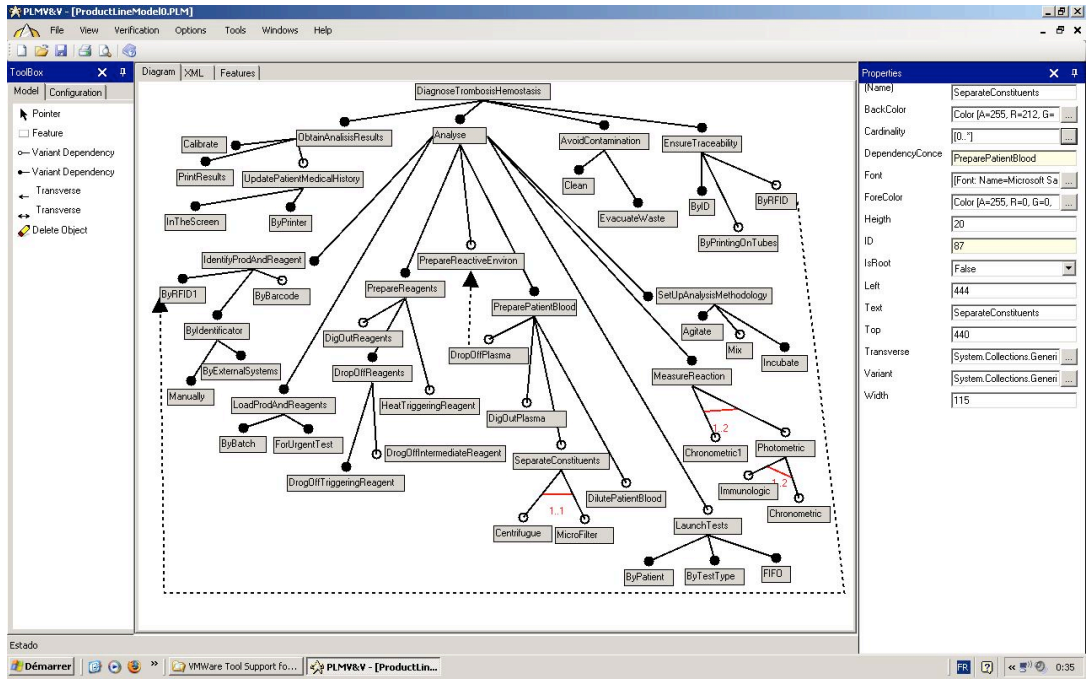

Figure 2. Extract of STAGO PL specification developed with our tool. Each view on the PL (FORE, XMI, textual constraints on features) can be accessed by a specific tab. Details on the specifications are given in [13].

Using FD constraints allowed us to specify the same constraints as the one that we had identified to reason about costs and revenue of each features. To do so, we associated [0..n] attributes to each features to specify costs and benefits. We, of course, had to define a fix value for n – we chose to use the same maximum cost and revenue for all feature for the purpose of the study.

For example, we specified constraints on the minimal number of measurement wells depending on the required tests and the required cadence for these tests.

*Chronometric.NumberOfWells +*
*Colorimetric.NumberOf Wells +*
*Immunologic.NumberOfWells >=*
*max(LaunchTest.TestCadence) ***
*max(LaunchTest.TestDuration)*

We could also specify that the initially optional function 'Agitate' must be implemented whenever one of the tests TP, TT, Fib, VwF or DDi are included

*(LaunchTest.TestType <> TCA) ∨ (LaunchTest.TestType*
*<> ATIII) ∨ ((LaunchTest.TestType <> PC) =>*
*Agitate = 1*

Looking at our list of FD specific constraints, we identified constraints that could not be specified before (ie with {true, false} features), namely:
- constraints on both [0..n] features and feature attributes. For example, we could play with the number of chronometric, colorimetric and immunologic measure dwells and specify a constraint on the number of their occurrence
- Chronometric + Colorimetric + Immunologic >= LaunchTest.TestCadence * LaunchTest.TestDuration
- symbolic constraints such as:
- Atmost (1, [Agitate, Mix, Incubate], 2]
- to specify that each activity in a methodology can be repeated at most twice
- Another example of use of symbolic constraints was to specify possible combinations of value of the cadence, duration, and kind of determination for different kinds of test types:

*Relation ([LauchTest.TestType, LauchTest.TestDuration,*
*LauchTest.TestCadence, determination], [*
    *[TP, 2, 14, simple],*
    *[TP, 2, 14, double],*
    *[TCA, 2, 14, simple],*
    *[TT, 3, 2, double],*
    *[Fib, 10, 5, double],*
    *[ATIII, 15, 3, double],*
    *[VwF, 13, 8, double],*
    *[PC, 2, 6, simple],*
    *[DDi, 6, 8, simple]])*

- Last, we were able to specify reified constraints such as

*LaunchTest.TestType = TCA <=> C*
*C => Chronometric = 1 ∧ Chronometric.Speed = normal*

which enforces the use chronometric measurement technique when TCA test is demanded. It specifies also the required speed for this test.

We also used feature attributes to support cost/benefit analysis on measurement techniques. The following goals could for instance be specified:

*Min (Chronometric.Cost **
*Chronometric.NumberOfWells + Colorimetric.Cost **

*Colorimetric.NumberOfWells + Immunologic.Cost \* Immunologic.NumberOfWells),* and

*Max (Chronometric.Revenue \* Chronometric.NumberOfWells + Colorimetric.Revenue \* Colorimetric.NumberOfWells + Immunologic.Revenue \* Immunologic.NumberOfWells)*

Our observations are also the following ones:

- Incremental development and maintenance of PL models is made possible as long as models are modified by adding constraints.
- GNU-Prolog computes very efficiently a first complete solution w.r.t. the selected/excluded features. In practice, this helped us in the configuration process as it provided a general idea of the product that was being built.
- GNU-Prolog computation of the next solution was effective as it offered an alternative to the configurations that had already been explored. Iterating over this function allowed to review the various solutions one by one – or to identify that the variability space was still very open by counting the number of remaining configurations that satisfied the constraints for the requirements at hand.

These results are encouraging and confirm that CP over FD is well suited to precisely model and efficiently configure PL.

## VI. RELATED WORKS

This paper is not the first to explore the use of constraint programming in the context of PL. Some proposals had been made to support automatic analysis of feature-based models in order to allow retrieving information.

The greatest number of works to automate features analysis is based on propositional logic. First-order logic is used to check validity of feature models [16] [17] [18] but also to reason about them [19] [2] [1]. Our approach belongs to a family of approaches that relies on constraints specification (in particular the integer domain) rather than on {true, false} features. The simple fact of replacing the {true, false} domain by [0..1] opens the door to kinds of constraints that did not exist in the aforementioned approaches.

Czarnecki's proposals of staged configuration, valued features, and feature attributes has created an opportunity to move from boolean to integer constraints specification. Benavides's works [5] and [13] have in particular shown how PL could be analyzed by specifying integer constraints on attributes associated with features. In Benavides's approach, features themselves still have a {true, false} domain.

Our approach goes a step further by exploring FD Constraint Programming. For example, it shows how to deal with [0..n] features, how to deal with staged configuration, and it uses the versatility of constraint programming to provide numerous types of constraints that were not proposed in the approaches referenced before.

## VII. PERSPECTIVE ON FUTURE WORKS

An important challenge for RE in PL development is ensuring the consistency of requirements and product line variability so as to allow configuring the right products. This challenge is even more difficult to meet as, in practice, requirements are mostly expressed using several languages [20].

Indeed, it is widely recognized that the complexity of current software development justifies the simultaneous use of several models to specify and communicate various views and aspects of a system with regards to the involved stakeholders (executives, developers, distributors, marketing, architects, testers, etc.). These models include requirements stating the capabilities of the PL, but also other kinds of requirements such as strategic goals, development constraints and component reuse restrictions that are of most interest when making decisions about product configurations. In the other hand, an insufficient consideration of all necessary views leads to a problem understanding lack and so the failure of software projects.

Besides, the perception of variability often depends on the organization and the area of expertise of the involved stakeholders. It is not unusual that different delivered models are expressed using heterogeneous variability notations within a single project development. For example, analysts may deliver a requirements model based on use cases describing a high-level, user-oriented view on system functionality; while architects may deliver a feature-based model focusing on system structure and interaction from a more technical, design-oriented point of view. In the absence of a global view, and given the complexity of the communication, it's not surprising that requirements get missed or misunderstood.

What is clearly needed in the future is a systematic way to capture all the information given by the various viewpoints and to organize it so that missing information is more easily identified, the full impact of change is more easily understood, and dependencies are explicitly discerned so that configuration is facilitated. It would be beneficial as well if the captured information could refer to the same representation of variability so that models are better construed and integrated.

In this regard, a constraint programming based approach could easily be used for integrating PL models, then verifying their consistency, and configuring product lines in an integrated way, ie dealing with requirements from different viewpoints and various kinds of variable artifacts at the same time.

This paper only deals with constraint programming over FD. The outcome of our approach is multiple: first, constraints from different modeling languages can be specified and reasoned about. Second, constraints from different product line models can be integrated in a unique program and solved in an integrated way -even when the models are specified with different languages. Third, constraints that to our knowledge could not be specified in existing languages can be specified.

We explored constraint programming on finite domains, but many other domains could be relevant : Reals, Intervals, Rationals, Booleans, Rational Trees, Lists, Strings and Sets. Constraint Programming is versatile in that it adapts quite well to different applications. We have little doubt that the systematic exploration of these domains will generate new knowledge about product lines engineering.